\documentclass[twocolumn,epjc3]{svjour3}          

\RequirePackage[T1]{fontenc}

\smartqed  

\RequirePackage{graphicx}
\RequirePackage{mathptmx}      
\RequirePackage{flushend}
\RequirePackage[numbers,sort&compress]{natbib}
\RequirePackage[colorlinks,citecolor=blue,urlcolor=blue,linkcolor=blue]{hyperref}
\usepackage{amsmath}
\usepackage{overpic}
\journalname{Eur. Phys. J. C}

\begin{document}

\title{Feasibility study of measuring $b\to s\gamma$ photon polarisation in $D^0\rightarrow K_1(1270)^- e^+\nu_e$ at STCF}


\author{Yu-Lan Fan\thanksref{e1,addr1}
        \and
        Xiao-Dong Shi\thanksref{addr2,addr3} 
	Xiao-Rong Zhou\thanksref{e3,addr2,addr3}
	Liang Sun\thanksref{e2,addr1}
}

\thankstext{e1}{e-mail: \email {yulanf@whu.edu.cn}}
\thankstext{e2}{e-mail: \email {sunl@whu.edu.cn} 
(corresponding author)}
\thankstext{e3}{e-mail: \email {zxrong@ustc.edu.cn} 
(corresponding author)}

\institute{School of Physics and Technology, Wuhan University, Wuhan 430072, People's Republic of China\label{addr1}
          \and
          State Key Laboratory of Particle Detection and Electronics, Hefei 230026, People's Republic of China\label{addr2}
          \and
	  School of Physical Sciences, University of Science and Technology of China, Hefei 230026, People's Republic of China\label{addr3}
}

\date{Received: date / Accepted: date}

\maketitle

\begin{abstract}
We report a sensitive study of measuring $b\to s\gamma$ photon polarisation in $D^{0}\to K_1(1270)^-e^+\nu_e$ with an integrated luminosity of $\mathcal L$ = 1 ab$^{-1}$ at a center-of-mass energy of 3.773~GeV at a future Super Tau Charm Facility. 
More than 61,000 signals of $D^{0}\to K_1(1270)^-e^+\nu_e$ are expected.
	Based on a fast simulation software package, the statistical sensitivity for the ratio of up-down asymmetry is estimated to be $1.5\times 10^{-2}$ by performing a two-dimensional angular analysis in $D^{0}\to K_1(1270)^-e^+\nu_e$.
Combining with measurements of up-down asymmetry in $B\to K_1\gamma$, the photon polarisation in $b\to s\gamma$ can be determined model-independently.
\end{abstract}

\section{Introduction}

The new physics (NP) and related phenomena beyond the Standard Model (SM) could be explored by indirect searches in $b\to s\gamma$ processes.
The photon emitted from the electroweak penguin loop in $b\to s\gamma$ transitions is predominantly polarised in SM.
New sources of chirality breaking can modify the $b\to s\gamma$ transition strongly as suggested in several theories beyond the SM~\cite{papCPABDecay,papNPBDecay,papWisonCoef}.
A representative example is the left-right symmetric model (LRSM)~\cite{papLRSMYFSH,papNPYFSH}, in which the photon can acquire a significant right-handed component.
An observation of right-handed photon helicity would be a clear indication for NP~\cite{papLRSMYFSH}.

The effective Hamiltonian of $b\to s\gamma$ is
\begin{equation}
\label{hamilton}
        \mathcal {H}_{eff} = -\frac{4G_{F}}{\sqrt{2}}V_{tb}V^*_{ts}(C_{7L}\mathcal{O}_{7L}+C_{7R}\mathcal{O}_{7R}),
\end{equation}
where $C_{7L}$ and $C_{7R}$ are the Wilson coefficients for left- and right-handed photons, respectively. 
In the SM, the chiral structure of $W^{\pm}$ couplings to quarks leads to a dominant polarisation photon and a suppressed right-handed configuration, and the photon from radiative $\bar{B}$ ($B$) decays is predominantly left- (right-) handed, {\it i.e.}, $|C_{7L}|^2 \gg |C_{7R}|^2$ ($|C_{7L}|^2 \ll |C_{7R}|^2$).

Various methods have been proposed to determine the photon polarisation of the $b\to s\gamma$.
The first method~\cite{papWithBsToPsiGamma} suggests that the $CP$ asymmetries which depend on the photon helicity could be measured in time dependent asymmetry in the charged and neutral $B(t)\to$X$^{CP}_{s/d}\gamma$ decays.
The second method~\cite{papNPlepton} to determine the photon polarisation is based on $b\to s ~l^+ l^-$ transition where the dilepton pair originates from a virtual photon.
The third method in $\Lambda_b\to \Lambda\gamma$ also could be used to measure the photon polarisation directly.
The forward-backward asymmetry defined in~\cite{papLambdabToLambdaGam,papLambdabAtZ} is proportional to the photon polarisation.

Measuring the photon polarisation in radiative $B$ decays into $K$ resonance states, $K_{\rm res}(\to K\pi \pi)$,  is proposed in~\cite{papPhotonPolarKPiPiGam,papPhotonPolarBDecay}.
The photon polarisation parameter $\lambda_{\gamma}$ could be described by an up-down asymmetry ($A_{\rm UD}$) of the photon momentum relative to the $K\pi\pi$ decay plane in $K_{\rm res}$ rest frame.
The photon polarisation in $B\to K_{\rm res}\gamma$ is given in terms of Wilson coefficients~\cite{papPhotonPolarBDecay}:

\begin{equation}
\label{photon_Wilson}
        \mathcal \lambda_{\gamma} = \frac{|C_{7R}|^2-|C_{7L}|^2}{|C_{7R}|^2+|C_{7L}|^2},
\end{equation}
with $\lambda_{\gamma} \simeq -1$ for $b\to s \gamma$ and $\lambda_{\gamma} \simeq +1$ for $\bar{b}\to \bar{s} \gamma$.
The integrated up-down asymmetry which is proportional to photon polarisation parameter $\lambda_{\gamma}$ for the radiative process proceeding through a single resonance $K_{\rm res}$ is defined~\cite{papPhotonPolarKPiPiGam, papPhotonPolarBDecay}

\begin{equation}
\begin{aligned}
         {A}_{\rm UD} &= \frac{\Gamma_{K_{\rm res}\gamma}[\cos\theta_K>0]-\Gamma_{K_{\rm res}\gamma}[\cos\theta_K<0]}{\Gamma_{K_{\rm res}\gamma}[\cos\theta_K>0]+\Gamma_{K_{\rm res}\gamma}[\cos\theta_K<0]}\\
&= \lambda_{\gamma}\frac{3~{\rm Im}[\vec{n} \cdot (\vec{J}\times\vec{J^{*}})]}{4~|\vec{J}|^2}.
\end{aligned}
\end{equation}
where $\theta_K$ is defined as the relative angle between the normal direction $\vec{n}$ of the $K_{\rm res}$ decay plane and the opposite flight direction of the photon in the $K_{\rm res}$ rest frame, and $\vec{J}$ denotes the $K_{\rm res}\to K\pi\pi$ decay amplitude~\cite{papPhotonPolarKPiPiGam}.
In charm sector, radiative $D^0-$ decays into $CP$ eigenstate are expected to determine the photon polarization by means of the charm meson's finite width difference~\cite{papRadiactiveDPhotonPolarization}.
Recently, LHCb collaboration reported the direct observation of the photon polarisation with a significance of 5.2$\sigma$ in $B^+\to K^+\pi^-\pi^+\gamma$ decay~\cite{papPhotonPolarbTosGamLHCb}.
In the $K\pi\pi$ mass interval [1.1,1.3]~GeV/$c^2$ which is dominated by $K_1(1270)$, $A_{\rm UD}$ is extracted to be (6.9  $\pm$  1.7) $\times10^{-2}$.
However, the currently limited knowledge of the structure of the $K\pi\pi$ mass spectrum, which includes interfering kaon resonances, prevents the translation of a measured asymmetry into  an actual value for $\lambda_{\gamma}$. 

\begin{figure}[ht]
\centering
\includegraphics[width=1\linewidth]{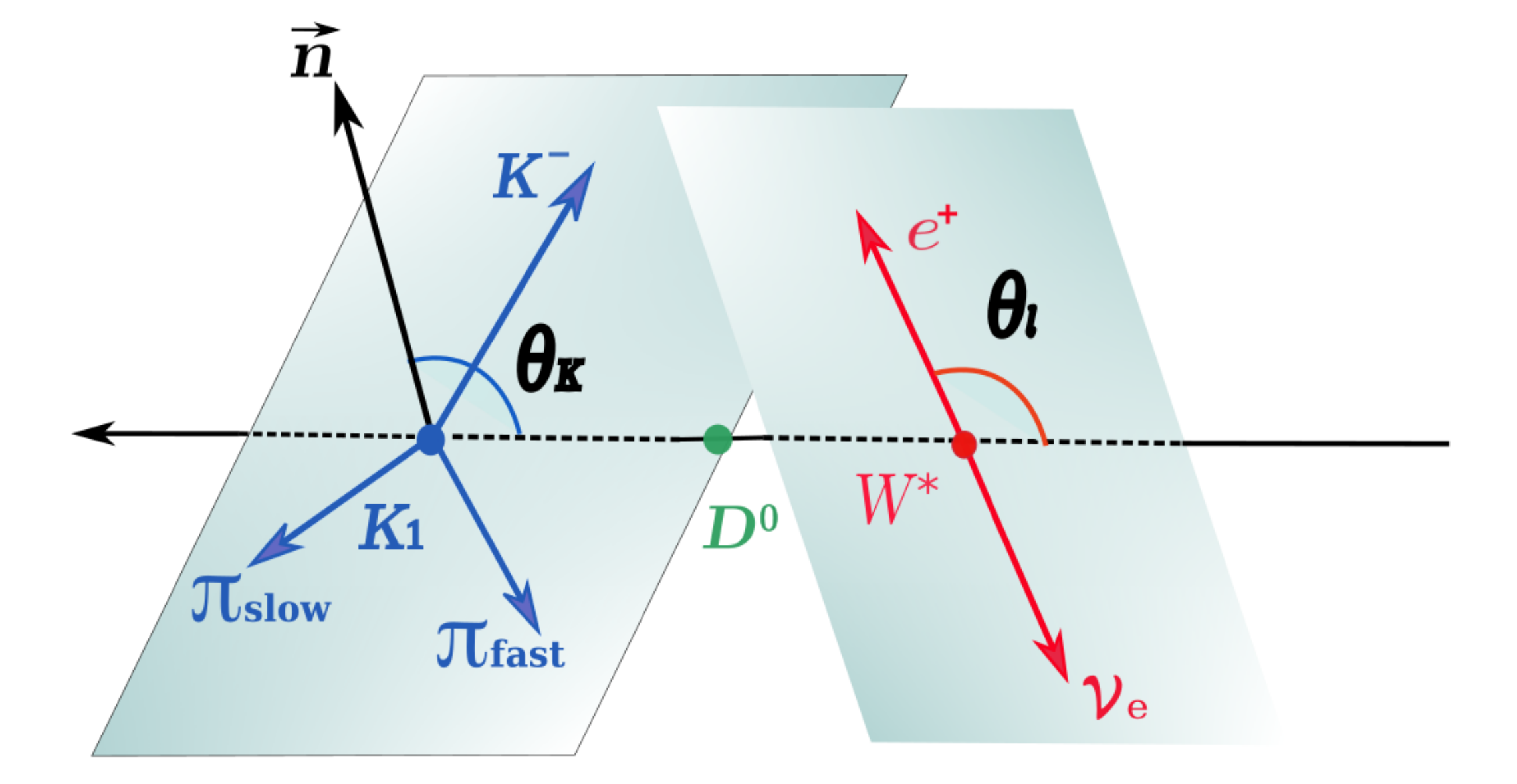}
	\caption{The kinematics for $D^0\to K_1^-(K^-\pi^+\pi^-) e^+\nu_e$. The relative angle between the normal direction of $K_1^-$ decay plane and the opposite of the $D^0$ flight direction in the $K_1^-$ rest frame is denoted as $\theta_K$, where the normal direction of $K_1^-$ decay plane is defined as $\vec{p}_{\pi,\rm slow}\times \vec{p}_{\pi,\rm fast}$ in which $\vec{p}_{\pi,\rm slow}$ and $\vec{p}_{\pi,\rm fast}$ corresponding to the momenta of the lower and higher momentum pions, respectively.  The $\theta_l$ is introduced as the relative angle between the flight direction of $e^+$ in the $e^+\nu_e$ rest frame and $e^+\nu_e$ in the $D^0$ rest frame.~\cite{papNovelMethod}}
\label{Angular}
\end{figure}

To solve this dilemma, 
three methods are proposed~\cite{papBtoJpsiKpipiAndBtoK1gamma, papNovelMethodDtoK1gamma, papNovelMethod}. In Ref.~\cite{papBtoJpsiKpipiAndBtoK1gamma}, by using $B\to J/\psi K_1\to J/\psi K\pi\pi$ channel, the hadronic information of $K\pi\pi$ can be determined. Along the lines of the method known to $B\to K_1(\to K\pi\pi)\gamma$ decays, the extraction of photon polarization in $D_(s)\to K_1(\to K\pi\pi)\gamma$ decays is introduced in the $K\pi\pi$ system~\cite{papNovelMethodDtoK1gamma}.
A novel method is proposed in Ref.~\cite{papNovelMethod} to determine the photon helicity in $b\to s\gamma$ by combining the $B\to K_1\gamma$ and semi-leptonic decay $D\to K_1l^+\nu_l(l=\mu^+,e^+)$ model-independently.
A ratio of up-down asymmetries is introduced in~\cite{papNovelMethod}.
Ref.~\cite{papNovelMethod} introduces two angles $\theta_K$ and $\theta_l$ in $D^0\to K_1^-e^+\nu_e$ shown in Figure~\ref{Angular} and the ratio of up-down asymmetry $A^{'}_{\rm UD}$ is defined as~\cite{papNovelMethod}

\begin{equation}
\begin{aligned}
        {A}^{'}_{\rm UD} &= \frac{\Gamma_{K_1^-e^+\nu_e}[\cos\theta_K>0]-\Gamma_{K_1^-e^+\nu_e}[\cos\theta_K<0]}{\Gamma_{K_1^-e^+\nu_e}[\cos\theta_l>0]-\Gamma_{K_1^-e^+\nu_e}[\cos\theta_l<0]}\\
&= \frac{{\rm Im}[\vec{n}\cdot(\vec{J}\times\vec{J^{*}})]}{|\vec{J}|^2}.
\end{aligned}
\label{Audp}
\end{equation}
Here the definition of the normal direction of $K_1$ decay plane is the same as in $B\to K_{\rm res}\gamma$ in LHCb~\cite{papPhotonPolarbTosGamLHCb}.

Then photon helicity parameter of $b\to s\gamma$ could be extracted by~\cite{papNovelMethod}

\begin{equation}
\label{photon_angle}
        \mathcal \lambda_{\gamma} = \frac{4~A_{\rm UD}}{3~A^{'}_{\rm UD}}.
\end{equation}
So the photon polarisation of $b\to s\gamma$ could be determined model-independently by the combination of $A_{\rm UD}^{'}$ in $D^0\to K_1^-(\to K^-\pi^+\pi^-)e^+\nu_e$ and $A_{\rm UD}$ in $B^+\to K_1^+(\to K^+\pi^-\pi^+)\gamma$.

Experimentally, the semileptonic decay of $D^0\to K_1(1270)^-$ $e^+ \nu_e$ has been observed for the first time with a statistical significance greater than 10$\sigma$ by using 2.93~fb$^{-1}$ of $e^+e^-$ collision data at $\sqrt{s}$ = 3.773~GeV by BESIII~\cite{papD0ToK1evBESIII}.
About 109 signals are observed, and the measured branching fraction is

\begin{equation}
\mathcal B(D^0\to K_1(1270)^- e^+\nu_e) = (1.09\pm0.13_{-0.16}^{+0.09}\pm0.12)\times 10^{-3}.\nonumber
\end{equation}
where the first and second uncertainties are the statistical and systematic uncertainties, respectively, and  the third uncertainty is the external uncertainty from the assumed branching fractions (BFs) of $K_1$ subdecays.

Still, the statistics of the current BESIII data set are insufficient to measure the ratio of up-down asymmetry in $D^0\to K_1(1270)^- e^+\nu_e$. 
A much larger data sample with similarly low background level is urgently needed for performing the angular analysis in $D^0\to K_1(1270)^- e^+\nu_e$, which calls for the construction of a next generation $e^+e^-$ collider operating at the $\tau$-charm energy region with much higher luminosity.

The Super Tau Charm Facility (STCF) is a scientific project proposed in China for high energy physics frontier~\cite{papSTCFCharm2018}.
The STCF plans to produce charmed hadron pairs near the charm threshold which allow for exclusive reconstruction of their decay products with well-determined kinematics.
Such samples at the threshold allow for a double-tag technique~\cite{papDTTechnique} to be employed where the full events can be reconstructed and provide a unique environment to measure $A^{'}_{\rm UD}$ in $D^0\to K_1(1270)^-e^+\nu_e$ with very low background level.

In this work, we present a feasibility study of a ratio of up-down asymmetry in $D^0\to K_1(1270)^-e^+\nu_e$ at STCF.
Throughout this paper, charged conjugated modes are always implied.
This paper is organised as follows:
In Sect.~\ref{section_detector}, detector concept for STCF is introduced as well as the Monte Carlo (MC) samples used in this feasibility study.
In Sect.~\ref{section_event}, the event selection and analysis method are described. The optimisation of detector response is elaborated in Sect.~\ref{section_optimization} and the results are presented in Sect.~\ref{section_statistical}. Finally, we conclude in Sect.~\ref{section_summary}.

\section{Detector and MC simulation}
\label{section_detector}
The proposed STCF is a symmetric electron-positron beam collider designed to provide $e^+e^-$ interactions at a center-of-mass (c.m.) energy $\sqrt{s}$ from 2.0 to 7.0~GeV.
The peaking luminosity is expected to be $0.5\times 10^{35}$ cm$^{-2}$s$^{-1}$ at $\sqrt{s}=$ 4.0~GeV, and the integrated luminosity per year is 1 ab$^{-1}$.
Such an environment will be an important low-background playground to test the SM and probe possible new physics beyond the SM.
The STCF detector is a general purpose detector designed for $e^+e^-$ collider which includes a tracking system composed of the inner and outer trackers,
a particle identification (PID) system with 3$\sigma$ charged $K/\pi$ separation up to 2~GeV/$c$,
and an electromagnetic calorimeter (EMC) with an excellent energy resolution and a good time resolution,
a super-conducting solenoid and a muon detector (MUD) that provides good charged $\pi/\mu$ separation. The detailed conceptual design for each sub-detector, the expected detection efficiency and resolution can be found in~\cite{papSTCFCharm2018, papSTCFIPAC2018, papFastSim}.

Currently, the STCF detector and the corresponding offline software system are under active development.
A reliable fast simulation tool for STCF has been developed~\cite{papFastSim}, which takes the most common event generators as input to perform a fast and realistic simulation.
The simulation includes resolution and efficiency responses for tracking of final state particles, PID system and kinematic fit related variables.
Besides, the fast simulation also provide some functions for adjusting performance of each sub-system which can be used to optimise the detector design according to physical requirement.

This study uses MC simulated samples corresponding to 1 ab$^{-1}$ of integrated luminosity at $\sqrt{s}$ = 3.773~GeV.
The simulation includes the beam-energy spread and initial-state radiation (ISR) in the $e^+e^-$ annihilations modeled with the generator {\sc kkmc}~\cite{papMCCalculation}.
The inclusive MC samples consist of the production of the $D\bar{D}$ pairs, the non-$D\bar{D}$ decays of the $\psi(3770)$,
the ISR production of the $J/\psi$ and $\psi(3686)$ states, and the continuum process incorporated in {\sc kkmc}~\cite{papMCCalculation}.
The known decay modes are modeled with {\sc evtgen~\cite{papBESEventGen}} using BFs taken from the Particle Data Group~\cite{papPDG2020},
and the remaining unknown decays from the charmonium states with {\sc lundcharm}~\cite{papEvtGenJpsi}.
Final-state radiation (FSR) from charged final-state particles is incorporated with the {\sc photos} package~\cite{papQED}.

Included in the inclusive $D\bar{D}$~MC sample, the $D^0$ $\to$ $K_1(1270)^-$ $e^+$$\nu_e$ decay is generated with the {\sc ISGW2} model~\cite{papISGW2} with BF comes from Ref.~\cite{papPDG2020} and $K_1(1270)^-$ meson is allowed to decay into all intermediate processes that result in a $K^-\pi^+\pi^-$ final state.
The resonance shape of the $K_1(1270)^-$ meson is parameterised by a relativistic Breit-Wigner function.
The mass and width of $K_1(1270)^-$ meson are fixed at the known values as shown in Table~\ref{K1_information},
and the BFs of $K_1(1270)$ subdecays measured by Belle~\cite{papKpipiBelle} are input to generate the signal MC events, 
since they give better consistency~\cite{papD0ToK1evBESIII} between data and MC simulation than those reported in~\cite{papPDG2020}. 

\begin{table}[htp]
\centering
	\caption{Mass, width~\cite{papPDG2020} and ratios of subdecays of $K_1(1270)^-$ (Fit2)~\cite{papKpipiBelle} used in this analysis.}
\label{K1_information}
\begin{tabular}{cc}
\hline
	Mass (GeV/$c^2$) & $1.253\pm0.007$\\
	Width (MeV) & $90\pm20$\\
\hline
	Decay mode & Decay ratio (\%)\\
\hline
	$K\rho$ &  $54.8 \pm 4.3\phantom{0}$\\
	$K_0^*(1430)\pi$ & $2.01 \pm 0.64$\\
	$K^*(892)\pi$ & $17.1 \pm 2.3\phantom{0}$\\
	$K\omega$ & $22.5 \pm 5.2\phantom{0}$ \\
\hline
\end{tabular}
\end{table}

\section{Event section and anasysis}
\label{section_event}
The feasibility study employs the $e^+e^-\to \psi(3770)\to D^0 \bar{D}^0$ decay chain.
The $\bar{D}^0$ mesons are reconstructed by three channels with low background level, $\bar{D}^0\to K^+\pi^-$, $K^+\pi^-\pi^0$ and $K^+\pi^-\pi^+\pi^-$.
These inclusively selected events are referred to as single-tag (ST) $\bar{D}^0$ mesons.
In the presence of the ST $D^0$ mesons, candidates for $D^0\to K_1(1270)^-e^+\nu_e$ are selected to form double-tag (DT) events.

Each charged track is required to satisfy the vertex requirement and detector acceptance in fast simulation.
The combined confidence levels under the positron, pion and kaon hypotheses ($CL_e$, $CL_{\pi}$ and $CL_{K}$, respectively) are calculated.
Kaon (pion) candidates are required to satisfy $CL_K > CL_{\pi}$ ($CL_{\pi} > CL_K$).
Positron candidates are required to satisfy $CL_e$ / ($CL_e$ + $CL_K$ + $CL_{\pi}$) $>$ 0.8.
To reduce the background from hadrons and muons, the positron candidate is further required to have a deposit energy in the EMC greater than 0.8 times its momentum in the MDC.
The $\pi^0$ meson is reconstructed via $\pi^0\to \gamma\gamma$ decay.
The $\gamma\gamma$ combination with an invariant mass in the range (0.115, 0.150)~GeV/$c^2$ are regarded as a $\pi^0$ candidates,
and a kinematic fit by constraining the $\gamma\gamma$ invariant mass to the $\pi^0$ nominal mass~\cite{papPDG2020} is performed to improve the mass resolution.
\begin{figure*}[t]
\centering
\begin{overpic}[width=5.5cm,angle=0]{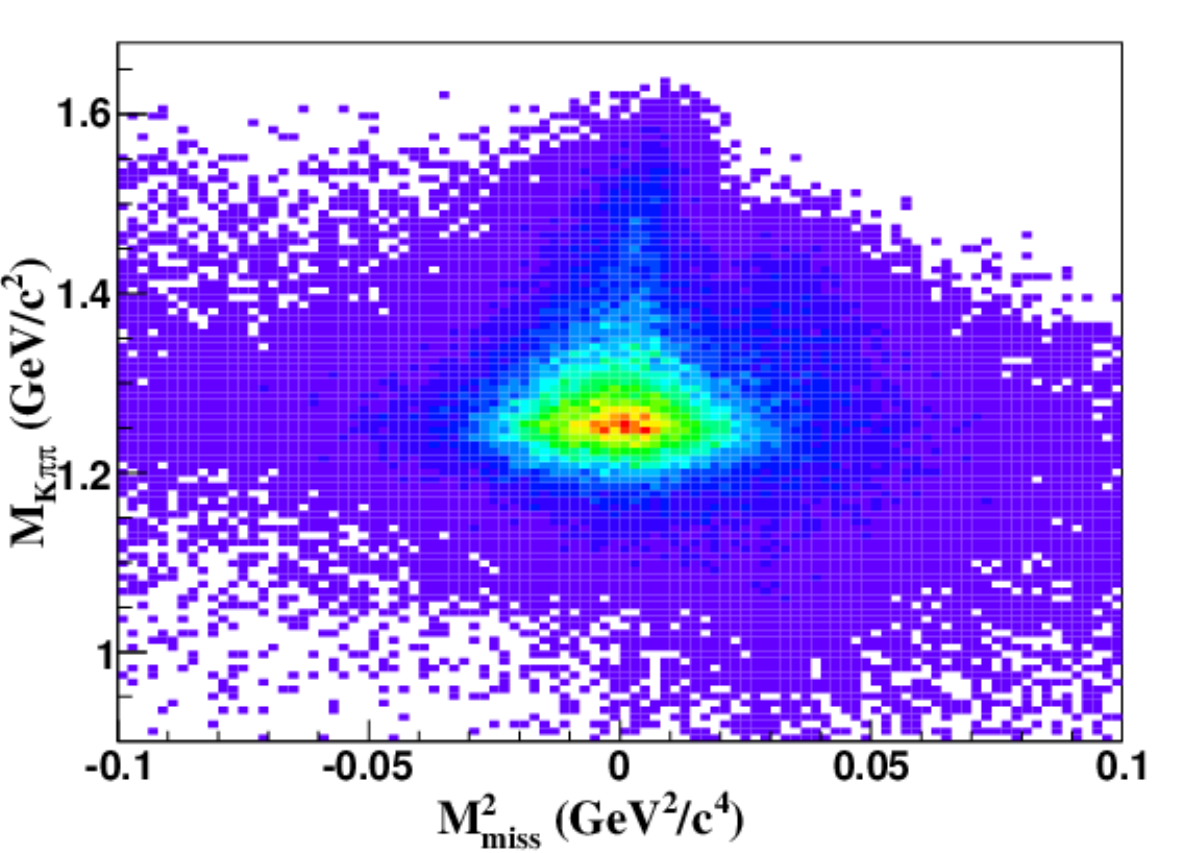}
\put(70,60){\textbf{(a)}}
\end{overpic}
\begin{overpic}[width=5.5cm,angle=0]{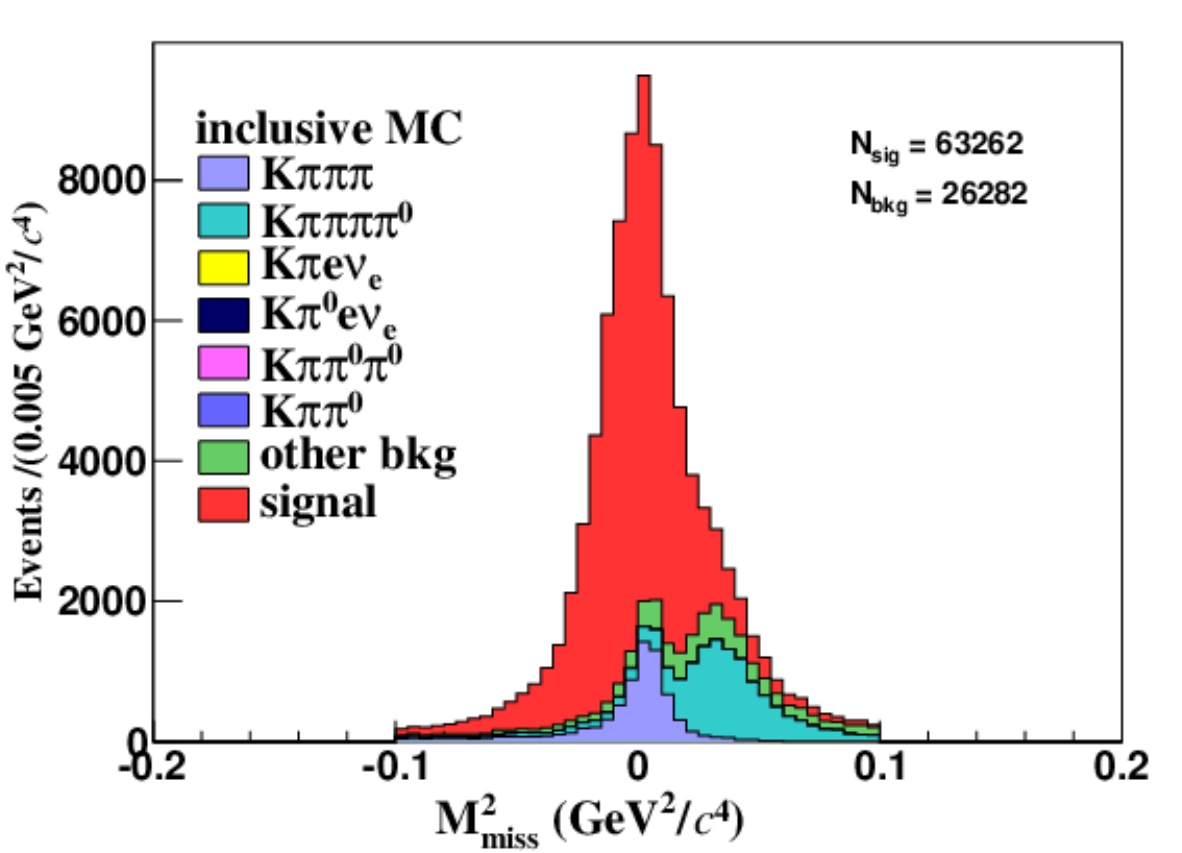}
\put(60,60){\textbf{(b)}}
\end{overpic}
\begin{overpic}[width=5.5cm,angle=0]{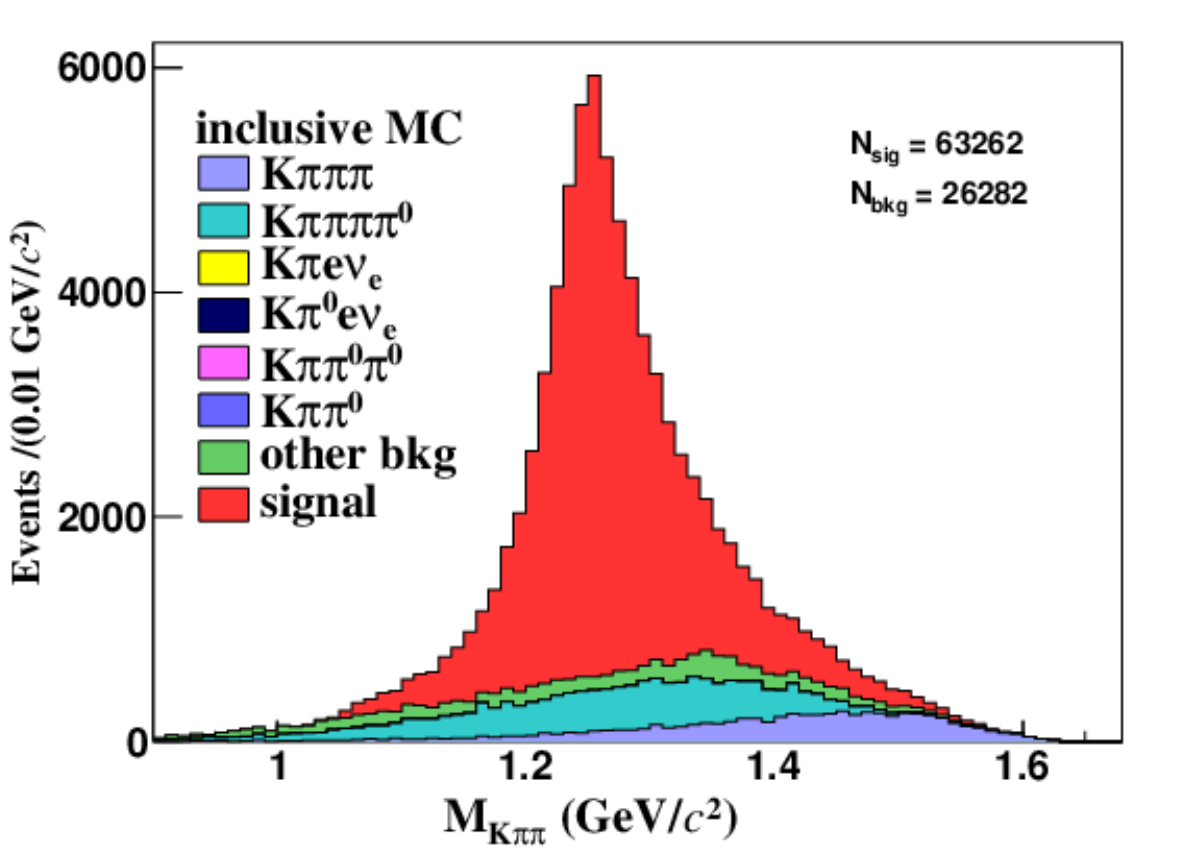}
\put(60,60){\textbf{(c)}}
\end{overpic}
\caption{(a).The $M_{\rm miss}^2$ vs. $M_{K\pi\pi}$ distribution of semi-leptonic candidate events. (b,c) the $M_{\rm miss}^2$/$M_{K\pi\pi}$ distribution of the semi-leptonic candidate events, where the red part denotes the signal events and other parts denote the remaining background events.}
\label{2Dhist}
\end{figure*}
The ST $\bar{D}^0$ mesons are identified by the energy difference $\Delta E \equiv E_{\bar {D}^0}-E_{\rm beam}$ and the beam-constrained mass $M_{\rm BC}$ $\equiv$ $\sqrt{E^2_{\rm beam}-|\vec{p}_{\bar{D}^0}|^2}$,
where $E_{\rm beam}$ is the beam energy, and $E_{\bar{D}^0}$ and $\vec{p}_{\bar{D}^0}$ are the total energy and momentum of the ST $\bar D^0$ in the $e^+e^-$ rest frame.
If there are multiple combinations in an event, the combination with the smallest $\Delta E$ is chosen for each tag mode. The combinatorial backgrounds in the $M_{\rm BC}$ distributions are suppressed by requiring $\Delta E$ within
(-29, 27), (-69, 38) and (-31, 28)~MeV for $\bar D^0\to K^+\pi^-$, $K^+\pi^-\pi^0$ and $K^+\pi^-\pi^+\pi^-$, respectively, which correspond to about 3.5$\sigma$ away from the fitted peak.

Particles recoiling against the ST $\bar D^0$ mesons candidates are used to reconstruct candidates for $D^0\to K_1(1270)^- e^+\nu_e$ decay, where the $K_1(1270)^-$ meson is reconstructed using its dominant decay $K_1(1270)^-\to K^-\pi^+\pi^-$.
It is required that there are only four good unused charged tracks available for this selection.
The charge of the lepton candidate is required to be the same as that of the charged kaon of the tag side.
The other three charged tracks are identified as a kaon and two pions, based on the same PID criteria used in the ST selection.
The kaon candidate must have charge opposite to that of the positron.

The main peaking background comes from misidentifying a pion to a positron,
and additional criteria as in~\cite{papD0ToK1evBESIII} are used to improve the $\pi$/$e$ separation.
Information concerning the undetectable neutrino inferred by the kinematic quantity $M^2_{\rm miss} \equiv E^2_{\rm miss} - |\vec p_{\rm miss}|^2$,
where $E_{\rm miss}$ and $\vec p_{\rm miss}$ are the missing energy and momentum of the signal candidate, respectively,
calculated by $E_{\rm miss} \equiv \vec E_{\rm beam} - \sum_{j}E_j$ and $\vec p_{\rm miss}\equiv -\vec p_{\bar{D}^0} - \sum_j\vec p_{j}$ in the $e^+e^-$ center-of-mass frame.
The index $j$ sums over the $K^-$, $\pi^+$, $\pi^-$ and $e^+$ of the signal candidate, and $E_j$ and $\vec p_j$ are the energy and momentum of the $j$-th particle, respectively.
To partially recover the energy lost to the FSR and bremsstrahlung, the four-momenta of photon(s) within 5$^\circ$ of the initial positron direction are added to the positron four-momentum.

\begin{figure}[hb]
\centering
\includegraphics[width=1\linewidth]{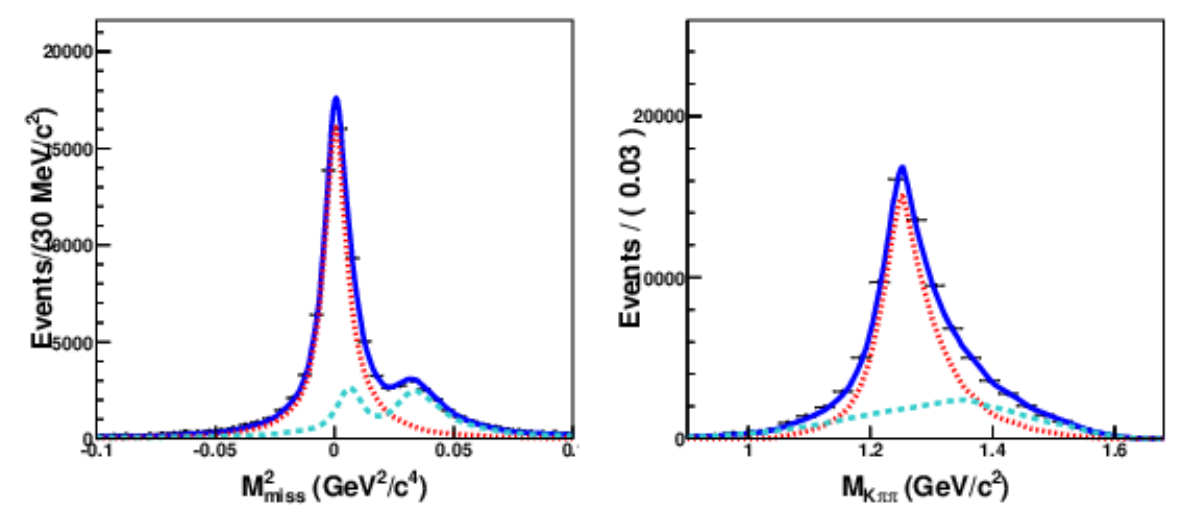}
\caption{Projection of the $M^2_{\rm miss}$(left) and $M_{K^-\pi^+\pi^-}$(right) of the DT candidate events of all three tag channels. The point with error bar are MC sample; the blue solid red dotted and green dashed curves are total fit, signal and background, respectively.}
\label{2Dfit}
\end{figure}

Figure~\ref{2Dhist} shows the distribution of $M_{K^-\pi^+\pi^-}$ vs. $M^2_{\rm miss}$ of the accepted $D^0\to K^-\pi^+\pi^-e^+\nu_e$ candidate events in the MC sample after combining all tag modes. 
A clear signal, which concentrates around the $K_1(1270)^-$ nominal mass in the $M_{K^-\pi^+\pi^-}$ distribution and around zero in the $M^2_{\rm miss}$ distribution, can be seen.
The selection efficiency of signal candidates with the ST modes $\bar {D}^0\to K^+\pi^-$, $K^+\pi^-\pi^0$ and  $K^+\pi^-$ $\pi^-\pi^+$ are 12.11$\%$, 6.93$\%$ and 6.25$\%$, respectively.
In order to determine the angular distributions of $\cos\theta_K$ and $\cos\theta_l$, a two-dimensional (2-D) fit to $M^2_{\rm miss}$ and $M_{K^-\pi^+\pi^-}$ is performed to extract the signal yield in each angle bin.
The 2-D fit projections to the $M^2_{\rm miss}$ and $M_{K^-\pi^+\pi^-}$ distributions are shown in Figure~\ref{2Dfit}.
In the fit, the 2-D signal shape is described by the MC-simulated shape extracted from the signal MC events while the 2-D background shape is modeled by those derived from the inclusive MC sample.
The smooth 2-D probability density functions of signal and background are modeled by using RooNDKeysPdf~\cite{papRooFit, papRooNDKeysClass}.

\begin{figure}[h]
\includegraphics[width=0.9\linewidth]{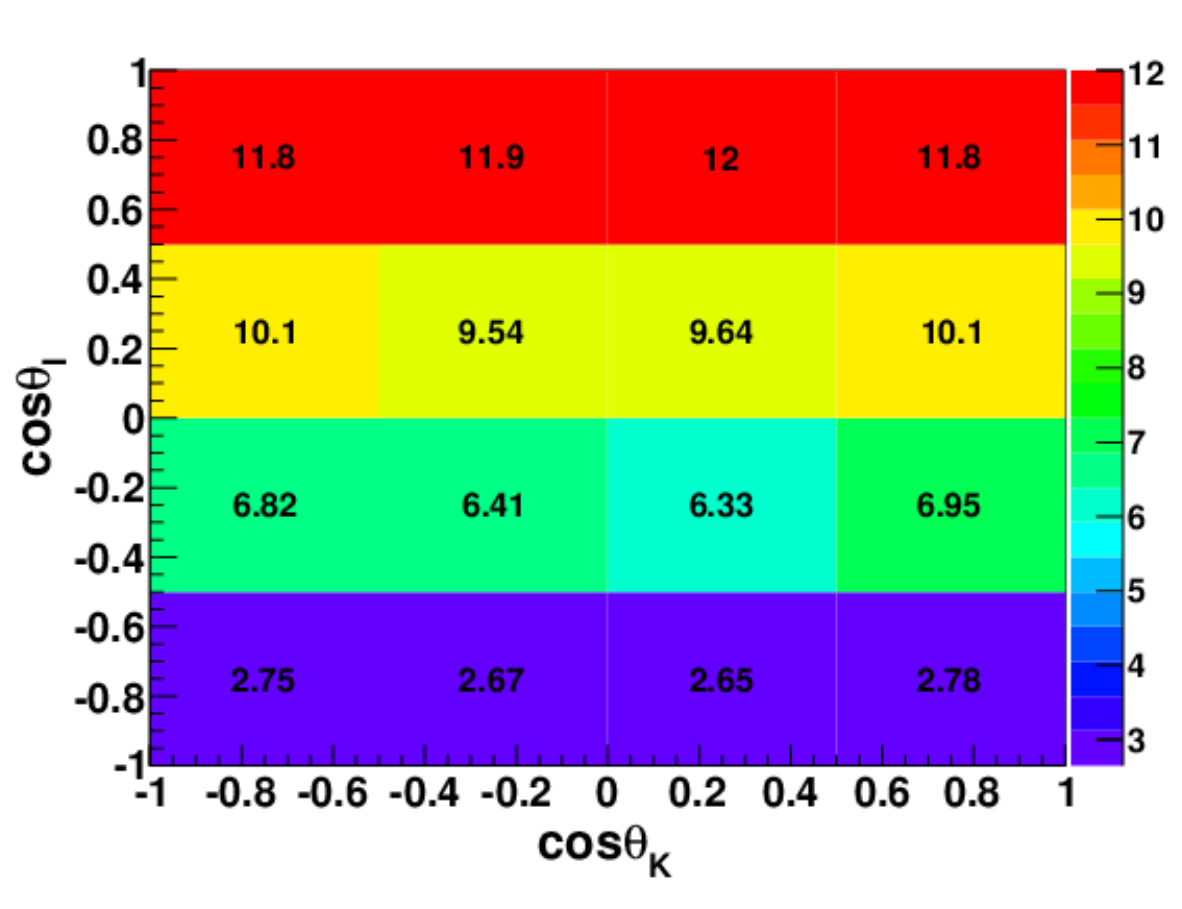}
	\caption{The signal reconstruction efficiencies (in percentage) in bins of $\cos\theta_K$ and $\cos\theta_l$. In each bin $j$, the signal reconstruction efficiency is obtained by $\epsilon_{\rm DT}^{j} = \frac{\sum_{i}\mathcal B_{\rm ST}^{i}\epsilon_{\rm DT}^{ij}}{\sum_{i}\mathcal B_{\rm ST}^i}$, where the $\mathcal B_{ST}^i$ denotes the known BF of each tag mode $i$, the $\epsilon_{DT}^{ij}$ represents the DT efficiency in each bin $j$ for tag mode $i$.}
\label{angular_eff}
\end{figure}

The reconstructed efficiencies of signal candidates in each $\cos\theta_K$ and $\cos\theta_l$ interval are shown in Figure~\ref{angular_eff}.
The signal reconstruction  efficiency shows a clear trend of increasing monotonically with $\cos\theta_l$, which is due to strong correlation between $\cos\theta_l$ and electron momentum, and $D^0$ candidates with lower momentum electrons are less likely to satisfy electron tracking and PID requirements.

The signal yields in each $\cos\theta_K$ and $\cos\theta_l$ interval corrected by the signal reconstruction efficiency are fitted with a polynomial function~\cite{papNovelMethod}

\begin{equation}
\begin{split}
f(\cos\theta_K, \cos\theta_l; & A_{\rm UD}^{'}, d_{+}, d_{-}) = \\
                &(4+d_+ + d_-) [1+\cos^2\theta_{K}\cos^2\theta_l]\\
                &+ 2(d_+-d_-) [1+\cos^2\theta_K]\cos\theta_l\\
                &+ 2A_{\rm UD}^{'}(d_+-d_-) \cos\theta_K[1+\cos^2\theta_l]\\
                &+ 4A_{\rm UD}^{'}(d_++d_-) \cos\theta_K\cos\theta_l\\
                &- (4-d_+-d_-) [\cos^2\theta_K+\cos^2\theta_l].
\end{split}
\label{polyfunc}
\end{equation}
where the $d_{\pm}$ are the angular coefficients, defined as:
\begin{equation}
\begin{aligned}
                d_+ =\frac{|c_+|^2}{|c_0|^2},
                d_- =\frac{|c_-|^2}{|c_0|^2}
\end{aligned}
\label{angular_coefficients}
\end{equation}
The coefficients $c_{\pm}$ and $c_0$ correspond to the nonperturbative amplitudes for $D$ decays into $K_1$ with transverse and longitudinal polarisations, respectively.
The ratio of up-down asymmetry $A_{\rm UD}^{'}$ can be extracted directly.
Besides, the fraction of longitudinal polarisation $\frac{|c_0|^2}{|c_0|^2+|c_+|^2+|c_-|^2}$ can be derived from the fitted $d_{\pm}$ values.
Form factor calculations based on different approaches such as covariant light-front quark model(LFQM) and light-cone QCD sum rules(LCSR) obtain the different results significantly.~\cite{papAudFFWangWei, papSemileptonicLCSR}.

\begin{figure}[b]
\includegraphics[width=1\linewidth]{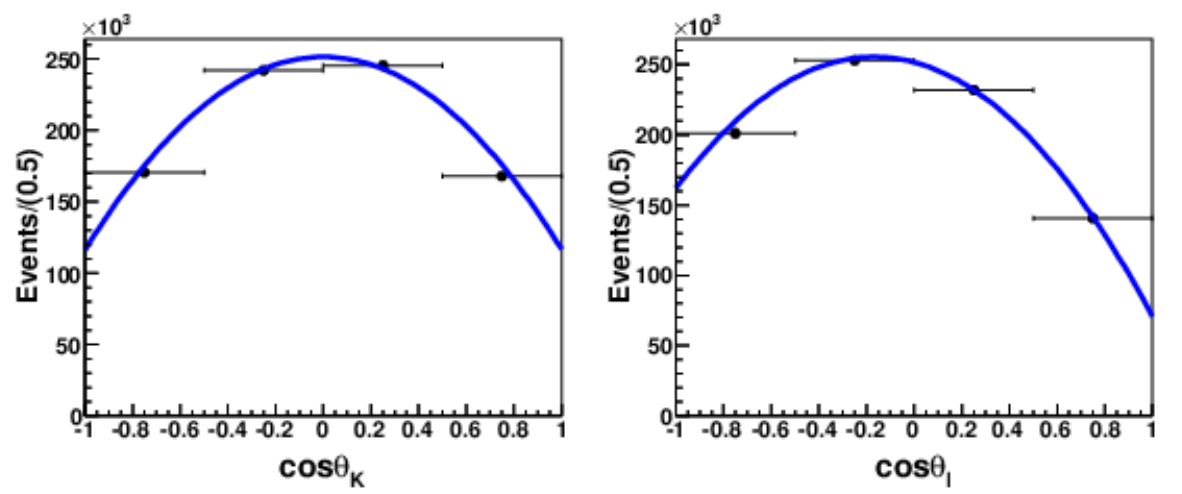}
\caption{Efficiency corrected signal yields in bins of $\cos\theta_K$(left) and $\cos\theta_l$(right). The curve is the result of fit using polynomial function.}\label{angular_fit}
\end{figure}

\begin{figure*}[ht]
\centering
\begin{overpic}[width=5.5cm,angle=0]{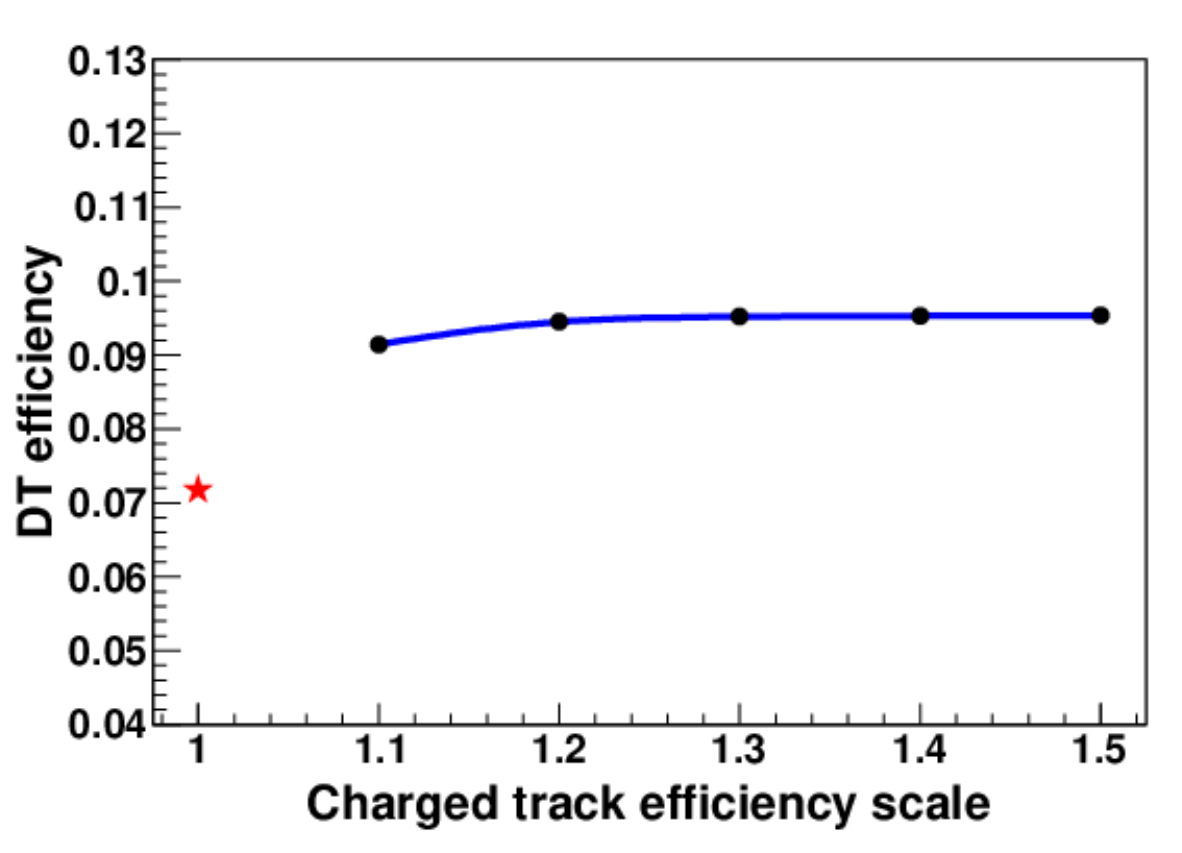}
\put(80,60){\textbf{(a)}}
\end{overpic}
\begin{overpic}[width=5.5cm,angle=0]{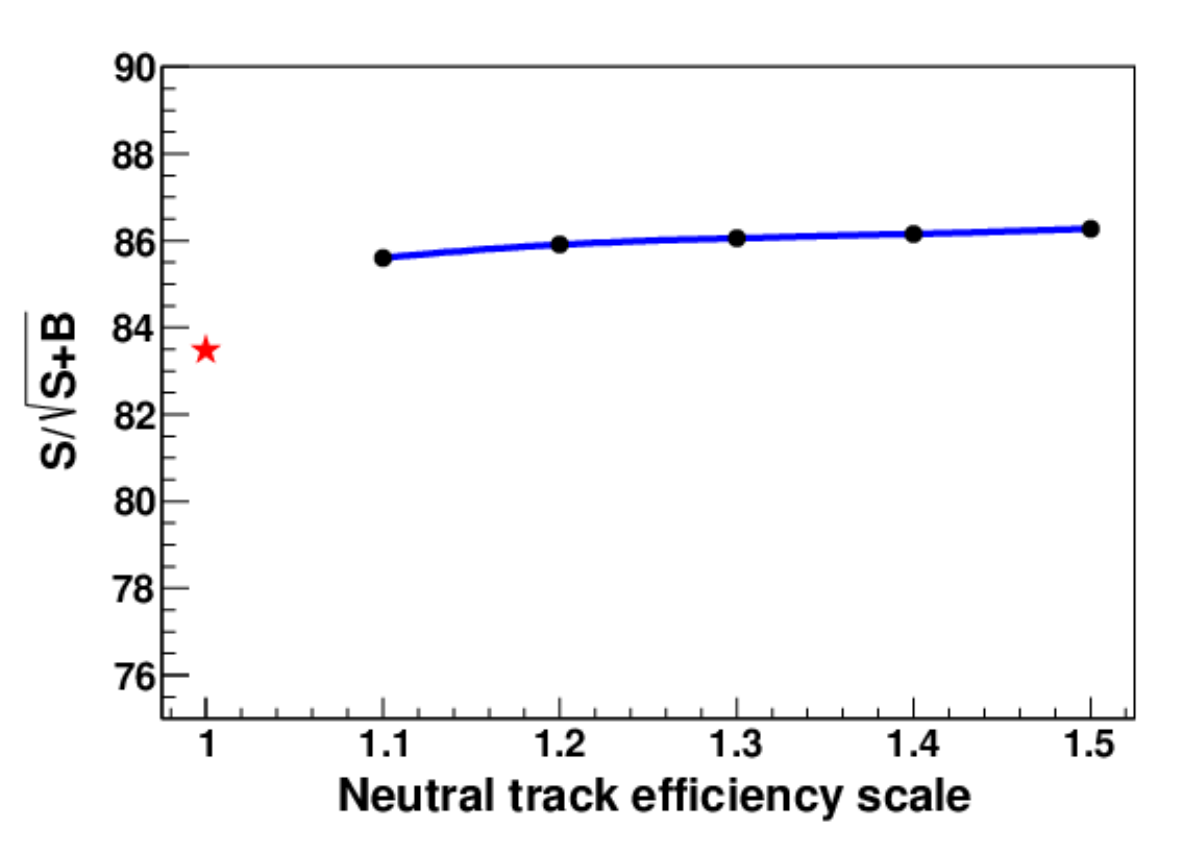}
\put(80,60){\textbf{(b)}}
\end{overpic}
\begin{overpic}[width=5.5cm,angle=0]{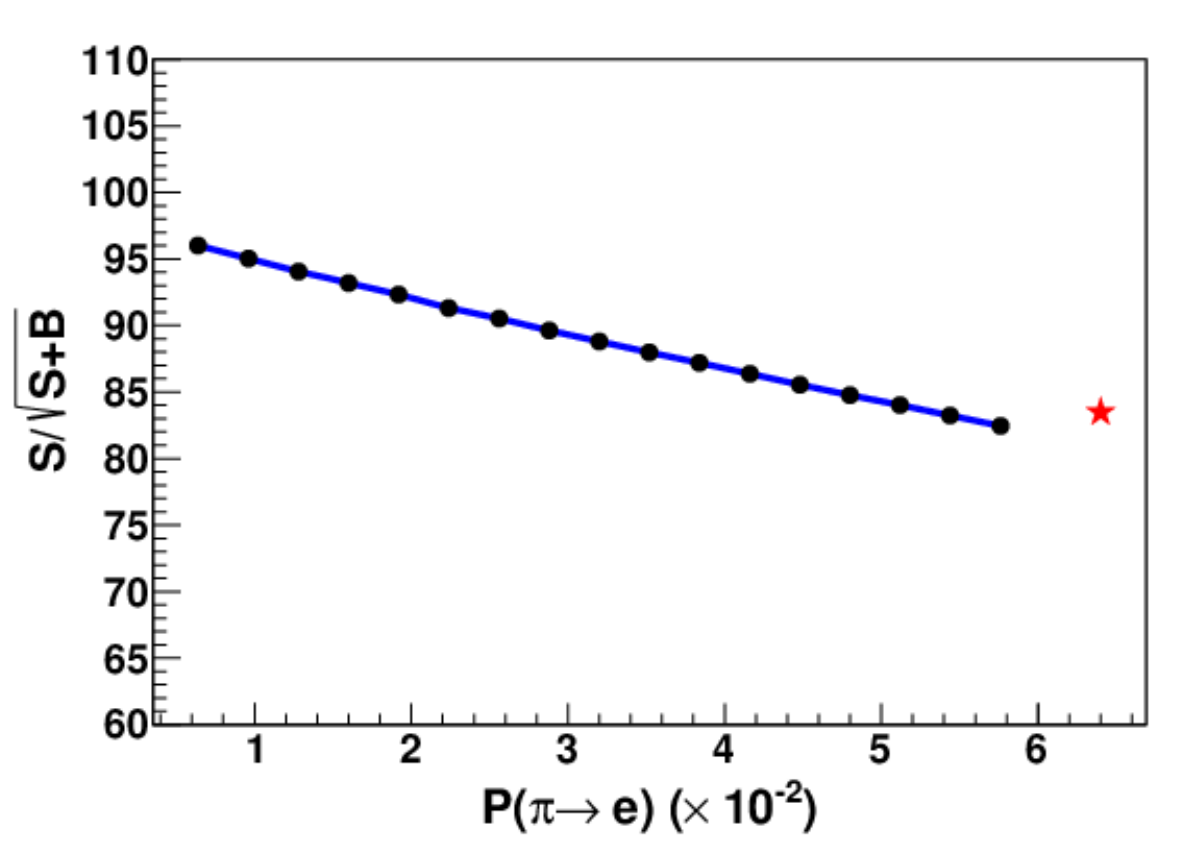}
\put(80,60){\textbf{(c)}}
\end{overpic}
\caption{The optimisation of DT efficiency for charged tracks of reconstructed efficiency (a); The optimisation of figure-of-merit for neutral tracks of reconstructed efficiency (b); The optimisation of figure-of-merit for misidentification from $\pi^+$ to $e^+$ (c). And the red star denotes the default result.}
\label{opt_abc}
\end{figure*}
The possibility of some events migrating from a bin to its neighbor caused by the detection resolution is considered by calculating the full width at half maximum(FWHM) of cos$\theta_{l}$ and cos$\theta_K$, respectively.
The value of FWHM is 0.115 and 0.05, which indicates that the bin migration effects can be ignored, due to the larger bin width of 0.5.

A 2-D $\chi^2$ fit to the $\cos\theta_l$ and $\cos\theta_K$ distributions allows to extract $A_{\rm UD}^{'}$, and the fit projections are shown in Figure~\ref{angular_fit}. The statistical sensitivity of $A_{\rm UD}^{'}$ based on 1 ab$^{-1}$ MC sample is thus determined to be in the order of 1.8$\times$10$^{-2}$.
As a cross-check, $A_{\rm UD}^{'}$ is also determined with a counting method according to Eq.~\ref{Audp} and the corresponding result is 
compatible with the angular fit method. However, the angular fit method yields a more precise result on $A_{\rm UD}^{'}$ and is taken as the nominal result.

\section{Optimization of detector response}
\label{section_optimization}

The main loss of the signal efficiency comes from the effects of charged tracking selection, neutral selection and identification of electron at low momentum.
These effects correspond to the sub-detectors of the tracking system, the EMC and the PID system.
By studying the DT efficiencies or signal-to-background ratios for this process with variation of the sub-detector's responses, the requirement of detector design can be optimised accordingly.
With the help of fast simulation software package, three kinds of detector responses are studied as introduced below:

~{\it a.Tracking efficiency} The tracking efficiency in fast simulation is characterised by two dimensions: transverse momentum $P_T$ and polar angle cos$\theta$, which are correlated with the level of track bending and hit positions of tracks in the tracker system.
For low-momentum tracks ($P_T$ < 0.2~GeV/c), it is difficult to reconstruct efficiently due to stronger electromagnetic multiple scattering,  electric field leakage, energy loss {\it etc.}.
However, with different technique in the tracking system design at STCF, or with advanced track finding algorithm, the efficiency is expected to be improved for low-momentum tracks.

Benefiting from the flexible approach to change the response of charged track, 
the detection efficiency is scaled with a factor from 1.1 to 1.5 in the fast simulation. 
The figure-of-merit, defined by DT efficiency for characterising the performance of tracking efficiency is shown in Figure~\ref{opt_abc}(a). From Figure~\ref{opt_abc}(a), it is found that the DT efficiency can be significantly improved with the given scale factors.
The resolution of momentum and position can also be optimised in fast simulation with proper functions.
Insignificant improvement is found among optimisation of absolute $\sigma_{xy}$ from 30 $\mu$m to 150 $\mu$m and absolute $\sigma_z$ from 500 $\mu$m to 2500 $\mu$m yet, where $\sigma_{xy}$ and $\sigma_z$ are the resolution of the tracking system in the $xy$ plane and $z$ direction, respectively.
This can be understood since the main source that affects the momentum resolution comes from electromagnetic multiple scattering on the material in the detector, instead of the position resolution. Therefore, material with low atomic number Z is required in the tracking system.

~{\it b.Detection efficiency for photon} In this analysis, $\pi^0$s are selected as part of the tag mode $\bar{D}^0\to K^+\pi^-\pi^0$ and the $\pi^0$ selection also helps to suppress the main background of $D^0\to K^-\pi^+\pi^-\pi^+\pi^0$ in signal side.
The figure-of-merit, defined by $\frac{S}{\sqrt{S+B}}$, to characterise the effect of photon detection efficiency on signal significance, in which $S$ denotes the expected signal yield of $D^0\to K_1(1270)^-(\to K^-\pi^+\pi^-) e^+\nu_e$ while $B$ denotes the background yield.
The value of $\frac{S}{\sqrt{S+B}}$ versus the scale factor of photon detection efficiency scanned from 1.1 to 1.5 is shown in Figure~\ref{opt_abc}(b). 

~{\it c.$\pi/e$ identification} Misidentification from a pion to electron in the momentum smaller than 0.6~GeV/$c$ forms the main peaking background $D^0\to K^-\pi^+\pi^-\pi^+$ and $D^0\to K^-\pi^+\pi^-\pi^+\pi^0$.
As the fast simulation provides the function for optimising the $\pi$/e identification which allows to vary the misidentification rate for $\pi$/$e$,
the $\pi$/$e$ misidentification rate at 0.2~GeV/$c$ is scanned from 5.7$\%$ to 0.64$\%$, shown in Figure~\ref{opt_abc}(c).
And $\frac{S}{\sqrt{S+B}}$ defined before is used to characterise the effect of misidentification of $\pi^+$ to $e^+$ on the signal significance.

In summary, three sets of optimization factors for different sub-detector responses are calculated separately: compared with our fast simulation with default settings,
the DT efficiency is improved by $\sim$27\% if the reconstructed efficiency for charged track is scaled by the factor of 1.1,
and the value of $\frac{S}{\sqrt{S+B}}$ is improved by 4\% if the photon detection efficiency is scaled by a factor of 1.1, or 7\% if the $\pi$/$e$ misidentification rate is lowered by half to 3.2\%, as reasonable assumptions in real case scenarios. 
With the above three factors applied altogether, the DT efficiency is improved by a factor of 33$\%$.
The corresponding angular 2-D $\chi^2$ fit based on updated efficiency-corrected signal yields  in different angular bins is performed. From the fit, the statistical uncertainty of the ratio of up-down asymmetry is extracted to be 1.5$\times 10^{-2}$, that is, 
improved by 17$\%$ compared with the no optimisation scenario.

\section{Statistical analysis}
\label{section_statistical}

With the above selection criteria and optimisation procedure, the 2-D simultaneous fit to $M_{\rm miss}^2$ vs. $M_{K\pi\pi}$ in the different interval of $\cos\theta_K$ vs. $\cos\theta_l$ is performed, the semi-leptonic decay signal yields produced are used for fitting the angular distribution.
Therefore, the sensitivity of the ratio of up-down asymmetry in $D^0\to K_1(1270)^-e^+\nu_e$ with an integrated luminosity of 1 ab$^{-1}$ is extracted as 1.5$\times$10$^{-2}$.
Besides, the selection efficiency for this process at $\sqrt{s}$ = 3.773~GeV where the cross section for $e^+e^-\to D^0\bar D^0$ = 3.6 nb~\cite{papCrossSection3770} is studied by a large MC sample,
with a negligible error.

\begin{figure}[ht]
\centering
\includegraphics[width=1\linewidth]{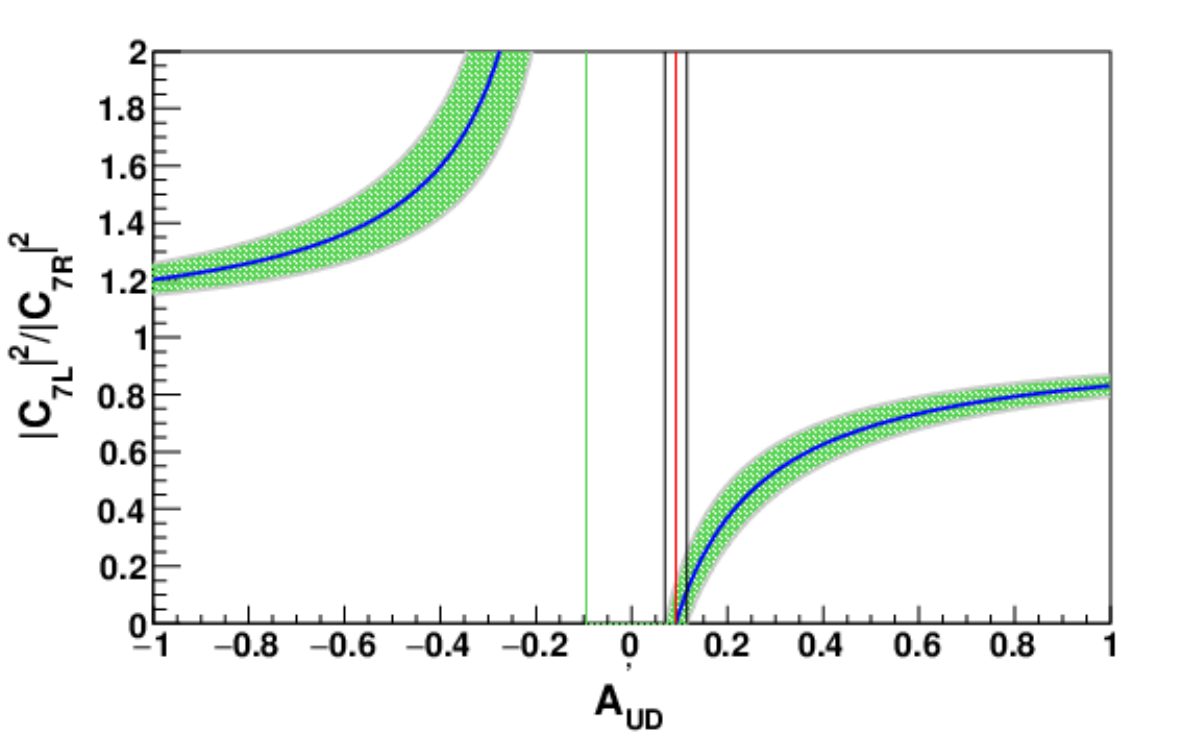}
\caption{Dependence of Wilson coefficient on ratio of up-down asymmetry, shown in the blue line, the green parts denote the consideration of uncertainties of $A_{\rm UD}$, the red solid line denotes $A_{\rm UD}^{'}$ corresponding to the photon polarisation parameter predicted in SM, with the consideration of uncertainties of $A_{\rm UD}$ shown between the black solid lines.}
\label{WilsonCo}
\end{figure}

Eq.~(\ref{photon_Wilson}) indicates that the Wilson coefficients can be constrained by measuring the uncertainty of the photon polarisation parameter $\lambda_{\gamma}$. Combining the uncertainty of $A_{\rm UD}$ measurement~\cite{papPhotonPolarbTosGamLHCb} with the uncertainty of $A_{\rm UD}^{'}$ measurement in this analysis, the sensitivity of $\lambda_{\gamma}$ can be determined using Eq.~(\ref{photon_angle}). Thus, the Wilson coefficients can be translated by the sensitivity of $A_{\rm UD}^{'}$.
Figure~\ref{WilsonCo} depicts the dependency of Wilson coefficients on ratio of $A_{\rm UD}^{'}$, using the $A_{\rm UD}$ measured in the $K\pi\pi$ mass range of (1.1,1.3)~GeV/$c^2$~\cite{papPhotonPolarbTosGamLHCb} as the input, shown in the blue solid line. Considering the uncertainty of $A_{\rm UD}$, the corresponding constraints shown in the green parts.
The photon polarisation parameter $\lambda_{\gamma}$ is predicted to be $\lambda_{\gamma} \simeq$ -1 for $\bar{b}\to \bar{s}\gamma$ in SM, which translated to $A_{\rm UD}^{'} \simeq$ (9.2 $\pm$ 2.3)$\times$10$^{-2}$ shown in the red and black solid line in Figure~\ref{WilsonCo}.

For the systematic uncertainty on $A_{\rm UD}^{'}$, possible sources include the electron tracking and PID efficiencies as functions of electron momentum which cannot cancel out in the $\cos\theta_l$ distribution due to strong correlation between $\cos\theta_l$ and electron momentum as mentioned before. 
With the current binning scheme as shown in Fig.~\ref{angular_eff}, the possibility of some events migrating from an angular bin to its neighbor because of the detector resolution effects on $\cos\theta_K$ and $\cos\theta_l$ is expected to be small and the related systematic uncertainty should be manageable.
Moreover, as in the BESIII analysis~\cite{papD0ToK1evBESIII}, the signal and background shape modeling would affect the signal yields considerably in different angular bins, due to imprecise knowledge on the $K_1(1270)$ line shape, and background events such as $D^0\to K^-\pi^+\pi^-\pi^+\pi^0$.

Our simulation does not include non-$K_1(1270)^-$ sources of $K^-\pi^+\pi^-$ in the $D^0 \to K^- \pi^+ \pi^- e^+ \nu_e$ decay, which are estimated to be at least one order of magnitude lower than our signal decay of $K_1(1270)^-$~\cite{papD0ToK1evBESIII}. We expect the systematic effect of the non-$K_1(1270)^-$ sources on $A^{'}_{\rm UD}$ to be small, although detailed studies on the $K_1(1400)^-$ contribution are needed when more data become available. 

\section{Summary and prospect}
\label{section_summary}

In this work, the statistical sensitivity of a ratio of up-down asymmetry $D^0\to K_1(1270)^-e^+\nu_e$ with an integrated luminosity of $\mathcal L$ = 1 ab$^{-1}$ at $\sqrt{s}$ = 3.773~GeV
and the optimised efficiency with the fast simulation, is determined to be 1.5$\times 10^{-2}$ by performing an angular analysis.
The hadronic effects in $K_1\to K\pi\pi$ can be quantified by $A_{\rm UD}^{'}$, therefore, combined with the measured up-down asymmetry $A_{\rm UD}$ in $B^+\to K_1^+(\to K^+\pi^-\pi^+)\gamma$~\cite{papPhotonPolarbTosGamLHCb}, the photon polarisation in $b\to s\gamma$ can be measured to probe the new physics.

\section{Acknowledgments}

The authors are grateful to Wei Wang, Fu-Sheng Yu, Hai-Long Ma and Xiang Pan for useful discussions.
We express our gratitude to the supercomputing center of USTC and Hefei Comprehensive National Science Center for their strong support.
This work is supported by the Double First-Class university  project foundation of USTC and the National Natural Science Foundation of China under Projects No. 11625523.


\end{document}